\documentclass[12pt]{iopart}

\usepackage{graphicx}
\usepackage{dcolumn}
\usepackage{bm}

\usepackage{multirow}
\usepackage{epsfig}
\usepackage{amssymb}
\usepackage{color}
\begin{document}

\title{Crossover between Silicene and Ultra-Thin Si Atomic Layers on Ag(111) Surfaces}
\author{Zhi-Xin Guo
\footnote{Present address: Xiangtan University, Xiangtan, Hunan 411105, China}
and Atsushi Oshiyama}
\address{Department of Applied Physics, The University of Tokyo, Tokyo 113-8656, Japan}

\begin{abstract}
We report on total-energy electronic structure calculations in the density-functional theory performed for the ultra-thin atomic layers of Si on Ag(111) surfaces.
We find several distinct stable silicene structures: $\sqrt{3}\times\sqrt{3}$, $3\times3$, $\sqrt{7}\times\sqrt{7}$ with the thickness of Si increasing from monolayer to quad-layer. The structural bistability and tristability of the multilayer silicene structures on Ag surfaces are obtained, where the calculated transition barriers infer the occurrence of the flip-flop motion at low temperature. The calculated STM images agree well with the experimental observations.
We also find the stable existence of $2\times1$ $\pi$-bonded chain and $7\times7$ dimer-adatom-stacking fault Si(111)-surface structures on Ag(111), which clearly shows the crossover of silicene-silicon structures for the multilayer Si on Ag surfaces.
We further find the absence of the Dirac states for multilayer silicene on Ag(111) due to the covalent interactions of silicene-Ag interface and Si-Si interlayer. Instead, we find a new state near Fermi level composed of $\pi$ orbitals locating on the surface layer of $\sqrt{3}\times\sqrt{3}$ multilayer silicene, which satisfies the hexagonal symmetry and exhibits the linear energy dispersion. By
examining
the electronic properties of $2\times1$ $\pi$-bonded chain structures, we find that the surface-related $\pi$ states of multilayer Si structures are robust on Ag surfaces.
\end{abstract}

\maketitle

\section{Introduction}
A honeycomb-structured two-dimensional (2D) atomic layer consisting of group IV atoms exhibits peculiar electronic properties due to a fact that electrons near the Fermi level ($E_{\rm F}$) follow effectively the massless Dirac equation
(Weyl equation) \cite{SW}. A well-known and only unequivocally measured example is graphene where intriguing properties such as an anomalous quantum Hall effect is observed \cite{geim,kim}. Another element in group IV, Si, which has sustained our modern life, should exhibit such fascinating properties \cite{takeda,cahangirov} thus opening a new door to the next-generation technology with its pronounced relativistic effects related to the spin degrees of freedom \cite{yao,ezawa1,ezawa2}.

The 2D structure of Si, known as silicene, is first synthesized on Ag(111) surfaces. The monolayer silicene on Ag(111) exhibits variety of structures. The superperiodicities of $4\times4$ \cite{vogt,lin,jamagotchian}, $\sqrt{13} \times \sqrt{13}$ \cite{lin,jamagotchian}, and $2\sqrt{3} \times 2\sqrt{3}$ \cite{jamagotchian,feng} with respect to $1\times1$ Ag(111) surface have been observed, and the simulated scanning tunneling microscopy (STM) images of theoretically determined structures reproduce the observed images excellently \cite{guo1,guo2}. Recently, the $\sqrt{3} \times \sqrt{3}$ silicene on Ag(111) with respect to $1\times1$ silicene has been reported by Chen {\it et al.}\cite{chen1,chen2}. They also claimed the flip-flop motion between the two stable hexagonal $\sqrt{3}$ structures to explain their honeycomb symmetry STM images at higher temperature and its freeze at lower temperature \cite{chen2}. However, our density-functional theory (DFT) calculations have clarified that Chen's observation should correspond to the bilayer silicene \cite{guo3}. On the other hand, the multilayer (including bilayer) silicene structures have been also synthesized on Ag(111) surfaces in
other experiments \cite{arafune,padova,resta,APL,2DM},
which interestingly exhibit the similar honeycomb $\sqrt{3}\times\sqrt{3}$ STM images as that reported by Chen {\it et al.}, irrespective of the thickness of silicene. Thus it is very important to clarify the structural details for the multilayer silicene on Ag(111).

On the other hand, Si surfaces are premier stages on which the fabrication of most electron devices is achieved. Structural characteristics of such Si surfaces are well identified experimentally and theoretically \cite{dabrowski}. In particular, the atomic layers of the (111) surface has the hexagonal network in the lateral plane, and thus the similarity and dissimilarity between the silicene and the surface atomic layers of Si(111) surface are intriguing. Actually the cleaved Si(111) surface shows $2 \times 1$ superperiodicity and then its annealing converts it to the $ 7 \times 7$ periodicity. The $2 \times 1$ structure has been identified as the $\pi$-bonded chain structure \cite{pandey} in which the top two Si layers are drastically reconstructed to form chains of $\pi$ orbitals associated with the five- and seven-membered rings of the first and the second surface layers. The $ 7 \times 7 $ structure is more complicated and a dimer-adatom-stacking-fault (DAS) model \cite{takayanagi,takayanagi2} is now established \cite{stich,brommer}. It is important and interesting to clarify how a crossover between the silicene structure and the reconstructed surface structure takes place in the multilayer silicene.

As for the electron states near $E_{\rm F}$, the situation is controversial for the monolayer silicene on Ag(111): Angular-resolved photoelectron spectroscopy (ARPES) measurements \cite{vogt,avila} show the existence of the electron state with the linear energy dispersion (Dirac state) for the $3\times3$ silicene on $4\times4$ Ag(111), whereas the DFT calculations clarify the absence of Dirac electrons due to the strong silicene-substrate interactions
\cite{guo1,guo2,wang,pflugradt,chenmx};
it is of note that no Landau-level sequences peculiar to Dirac electrons have been observed \cite{lin2}. From the interference patterns obtained by scanning tunneling spectroscopy (STS) measurements, Chen {\it et al.} claimed the presence of Dirac states with the linear dispersion \cite{chen1,chen2} for the $\sqrt{3}\times\sqrt{3}$ silicene on Ag(111), whereas Arafune {\it et al.} deduced the contrary conclusion, the absence of Dirac electrons, from essentially identical STS experiments \cite{chen-arafune}. Our DFT calculations show that the $\sqrt{3}\times\sqrt{3}$ silicene structure observed by Chen {\it et al.} actually correspond to the bilayer silicene and the linear energy dispersion is attributed to the $sp$ band of Ag \cite{guo3}.
Recently, the ARPES measurements also observed the linear energy dispersions for the $\sqrt{3}\times\sqrt{3}$ multilayer silicene on Ag(111), where the electron velocity is much smaller than that observed by Chen {\it et al.} The linear energy bands are recognized to be the Dirac $\pi$ and
anti-bonding
$\pi$ (labeled as $\pi^{*}$) bands of multilayer silicene \cite{padova,padova1}.
Details of the electronic structure of the multilayer silicene are still mysterious and further theoretical work is in great need for the clarification.

In this work, we report on total-energy electronic structure calculations for the ultra-thin atomic layers of Si on Ag(111) surfaces.
We find several distinct stable silicene structures: $\sqrt{3}\times\sqrt{3}$, $3\times3$, $\sqrt{7}\times\sqrt{7}$ with the thickness of Si increasing from monolayer to quad-layer. We also find the existence of $2\times1$ and $7\times7$ Si(111)-surface structures on Ag(111). The crossover of silicene-silicon structures on Ag surface appears when the thickness of Si layer becomes larger than 2.
We also find the structural bistability and tristability of the multilayer silicene structures, where the calculated transition barriers infer the existence of flip-flop motion at low temperature.
We further find the absence of the Dirac states for multilayer silicene on Ag(111) due to the covalent interactions of silicene-Ag interface and Si-Si interlayer. Interestingly, we find a new state near $E_{\rm F}$ composed of $\pi$ orbitals locating the top layer of $\sqrt{3}\times\sqrt{3}$ multilayer silicene, which satisfies the hexagonal
symmetry and exhibits the linear energy dispersion. Finally, we explore the electronic properties of Si(111)-surface structures on Ag surfaces.

\section{Calculations}
The total-energy electronic-structure calculations have been performed in the density functional theory using VASP code \cite{vasp1,vasp2}. As for the exchange-correlation functional, the vdW-DF \cite{dion,klimes}, being capable of treating the dispersion force is adopted \cite{ad1}.
Calculations for monolayer silicene and some of bilayer silicene have been also performed by the local density approximation (LDA) and the generalized gradient approximation (GGA) \cite{guo1,guo2}. The results obtained are essentially same among the three exchange-correlation functionals, although quantitative values for the total energies are different.
The electron-ion interaction is described by the projector augmented wave method \cite{paw}, and the cutoff energy of 250 eV in the plane-wave basis set is used. The Ag surface is simulated by a repeating slab model in which a five-atomic-layer slab is cleaved from the face-centered-cubic (fcc) Ag with the experimental lattice constant (4.09 {\AA}). The Ag slabs are separated from their images by the vacuum regions with the thickness of 14 {\AA}, 17 {\AA}, 21 {\AA}, and 24 {\AA} for the monolayer, bilayer, tri-layer  and quad-layer Si deposited on Ag(111), respectively. The geometry optimization is performed until the remaining forces become less than 0.02 eV/{\AA}. We have carefully examined the validity of the $k$-point mesh in Brillouin zone (BZ) integration, and found that the spacing between adjacent $k$-point being less than 0.016 \AA$^{-1}$ is sufficient to obtain converged results.

It has been recognized that the $3\times3$ and $\sqrt{7} \times \sqrt{7}$ silicene are commensurate to the $4\times4$ and $\sqrt{13}\times\sqrt{13}$ Ag(111), respectively \cite{guo1}.
To explore the $\pi$-bonded chain and DAS structures on Ag(111) surface, we prepare initial structures by depositing the Si layers that are cleaved from the Si(111) with the periodicity of the $2\times2$ (two times of the $2\times1$ unit cell) and $7\times7$ on the $\sqrt{7}\times\sqrt{7}$ and $2\sqrt{21} \times 2\sqrt{21}$ Ag(111), respectively, where the lattice mismatches are 0.4 \% and 1.4 \%.
Following the convention of the silicene layer, we define a single layer, i.e., a single bilayer, as consisting of 4 Si atoms at the upper positions and the remaining 4 Si atoms at the lower positions in the $2 \times 2$ lateral cell of the Si(111) surface.

To assess relative stability among various stable and metastable structures, we introduce the cohesive energy $E_{c}$ and the binding energy $E_{\rm b}$. They are defined as $E_{\rm c} = (E_{\rm Ag(111)} + N_{\rm Si} \mu_{\rm Si} - E_{\rm tot}) / N_{\rm Si}$ and $E_{\rm b} = (E_{\rm Ag(111)} + E_{\rm si-layers} - E_{\rm tot}) / N_{\rm Lsi}$, respectively. Here $E_{\rm tot}$, $E_{\rm Ag(111)}$ and $E_{\rm si-layers}$ are the total energies of the Si layers on Ag(111), the clean Ag(111) surface and the freestanding Si layers, respectively. $N_{\rm Si}$ and $N_{\rm Lsi}$ are the total number of Si atoms and the number of Si atoms per Si layer, respectively. $\mu_{\rm Si}$ is the chemical potential of Si which is adopted as the total energy of an isolated Si atom in this paper. The cohesive energy defined above is the energy gain to make Si ultra-thin layers on the Ag(111) surface from the clean Ag surface plus the constituent Si atoms. This may be a measure of the catalytic ability of the Ag surface in forming the Si ultra-thin layers. The binding energy is the measure of the energy gain to put the freestanding Si structures on the Ag(111) surface. In any case, $E_{\rm c}$ and $E_{\rm b}$ represent relative stability of various Si structures on the Ag(111) surface.

\section{Stabilities and structures of Si layers on Ag(111)}
In this section we present structural characteristics and energetics of ultra-thin Si atomic layers on Ag(111) surfaces. We examine Si monolayer (ML), bilayer (BL), tri-layer (TL) and quad-layer (QL) structures. An interesting finding is the wealth of the structural multi-stability being composed of a variety of distinct structures with the cohesive-energy difference in the range of less than 100 meV per Si atom (meV/Si). The $\pi$-bonded chain and DAS structures observed on the Si(111) surface are found to exist also in the ultra-thin Si layers on the Ag(111) surface.

\subsection{Monolayer Si on Ag(111)}
\begin{figure}
\begin{center}
\includegraphics[angle= 0, width=0.7\linewidth]{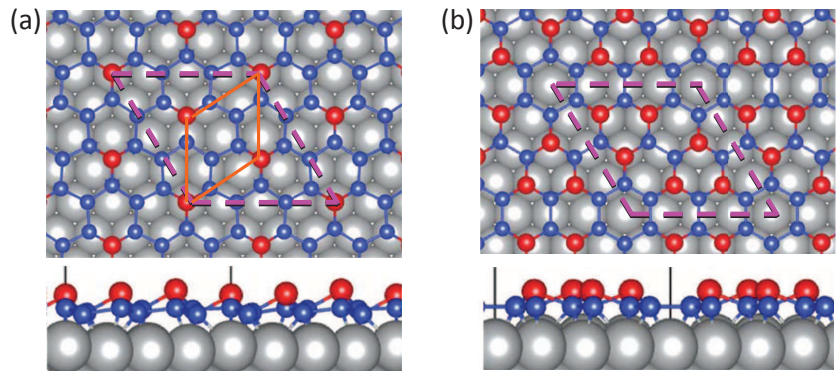}
\caption{
(color online)
Top (upper panels) and side (under panels) views of geometry optimized structures for the $3\times3$ monolayer (ML) silicene on $4\times4$ Ag(111). After optimization, the silicene shows the $\sqrt{3} \times \sqrt{3}$ (a) and the $3 \times 3$ (b) periodicities, respectively. The Si atoms in the top and bottom vertical positions are depicted by the large red and small blue balls, respectively. The large gray balls depict the positions of the substrate Ag atoms. The simulated lateral unit cell is indicated by the dashed (pink) lines in the top view, and its boundary is indicated by the solid (black) lines in the side views. The obtained rhombic $\sqrt{3}\times\sqrt{3}$ periodicity is indicated by the solid (orange) lines in (a).
}
\label{strSL}
\end{center}
\end{figure}

\begin{table}
\caption{
Calculated cohesive energy $E_{\rm c}$ (eV/Si), binding energy $E_{\rm b}$ (eV/Si) and structural parameters for the silicene layers and the surface layers of Si(111) on Ag(111). $d_{\rm si-Ag}$ ({\AA}) is the spacing between the bottom Si layer and the topmost Ag layer. Note that the Si layer is buckled. $\Delta z$ ({\AA}) represents the averaged amount of the buckling of the top Si layer and $d_{\rm si-si}$ ({\AA}) is the interlayer distance defined as the minimum spacing between top Si layer and its underlying Si layer.}
\label{str}
\begin{center}
\begin{tabular}{cccccccc}
  \hline
  \hline
            &~            &$E_{\rm c}$  &$E_{\rm b}$  & $d_{\rm si-Ag}$   & $d_{\rm si-si}$   & $\Delta z$  \\
  \hline
  ML        & $\sqrt{3}\times\sqrt{3}$     & 5.387       & 0.63        & 2.37        & ---        & 1.05        \\
            & $3\times3$                   & 5.410       & 0.65        & 2.17        & ---        & 0.86        \\
                                             \\
 BL         & $\sqrt{3}\times\sqrt{3}$ & 5.339       & 0.78        & 2.25         & 2.58       & 0.96       \\
            & $3\times3$-AA                & 5.342       & 0.79        & 2.18         & 2.47       & 0.81           \\
            & $\sqrt{7}\times\sqrt{7}$-AA$^{\prime}$ & 5.331       & 0.80        &2.27          &2.39        &0.99        \\
            & $2\times1$                   & 5.350       & 0.58         &2.26         &2.25        &0.98         \\
            & $7\times7$                   & 5.316       & 0.63         &2.23         &1.40        &---         \\
                                                         \\
  TL        & $\sqrt{3}\times\sqrt{3}$  & 5.384       &0.83      &2.26          &2.53          &1.08         \\
            & $3\times3$-AAA          & 5.364       &0.77      &2.18          &2.48          &0.89         \\
            & $3\times 3$-ABC$^{\prime}$              & 5.391       &0.85      &2.26          &2.41          &1.06         \\
            & $\sqrt{7}\times\sqrt{7}$-AAA$^{\prime}$ & 5.370       &0.80      &2.26          &2.40          &1.10         \\
            & $2\times1$                   & 5.403       & 0.81     &2.28          &2.24          &1.05         \\
                                                       \\
  QL         &$\sqrt{3}\times\sqrt{3}$  & 5.418       &0.86          &2.28      &2.52          &1.09         \\
            & $3\times 3$-AAAA             & 5.384       &0.73          &2.18      &2.49          &0.88         \\
            & $3\times 3$-ABCA$^{\prime}$             & 5.423       &0.88          &2.26      &2.44          &1.17         \\
            &$\sqrt{7}\times\sqrt{7}$-ABCA & 5.404       &0.89          &2.26      &2.49          &0.95         \\
            &$2\times1$                    & 5.434       &0.88          &2.27      &2.24          &1.06        \\
  \hline
  \hline
\end{tabular}
\end{center}
\end{table}

Structures of ML silicene have been explored by our previous DFT calculations \cite{guo1,guo2}. In this subsection, we present those results and then add more details. We have found two distinct stable structures with $3 \times 3$ silicene periodicity, and four distinct stable structures with $\sqrt{7} \times \sqrt{7}$ periodicity. The calculated cohesive energies of all the six structures are in the range of 5.906 - 5.975 eV in LDA, inferring the possible observation of several distinct structures.
Actually the calculated STM images of the three most stable structures agree with the experimental images \cite{vogt,lin,feng} excellently and two of the three metastable structures have been recently observed in experiments \cite{jia}.
The most stable $ 3 \times 3$ silicene obtained by the present vdW-DF calculation is shown in Fig.~\ref{strSL}(b), where 6 of 18 Si atoms in a unit cell protrudes by 0.86 {\AA}, whereas the remaining 12 Si keep nearly the same height in the lower positions. The structural difference between the LDA and the vdW-DF is tiny. The cohesive energy obtained by vdW-DF is 5.410 eV (Table \ref{str}), which is between the LDA value (5.972 eV) and the GGA value (5.215 eV).
We have also found a $\sqrt{3} \times \sqrt{3}$ ML silicene structure [Fig.~\ref{strSL}(a)] for the $3\times3$ silicene/$4\times4$ Ag(111) commensuration, in which one of six Si atoms in a unit cell protrudes about 1 {\AA}, whereas the remaining five Si keep nearly the same height in the bottom layer. Chen {\it et al.} \cite{chen2} have also found two distinct $\sqrt{3} \times \sqrt{3}$ structures both of which are different from the structure in Fig.~\ref{strSL}(a). We have clarified that the structures obtained by Chen {\it et al.} emerge when we perform the computation with insufficient k-point sampling \cite{guo3}.

The structural parameters for the stable $\sqrt{3}\times\sqrt{3}$ [Fig. \ref{strSL}(a)] and $3\times3$ [Fig. \ref{strSL}(b)] silicene on $4\times4$ Ag(111) are shown in Table \ref{str} along with the cohesive energies $E_{c}$ and the binding energies $E_{\rm b}$.
From Table \ref{str}, the $\sqrt{3}\times\sqrt{3}$ structure is 23 meV/Si less stable than the $3\times3$ silicene on Ag(111) which has been widely observed in experiments.
This shows that the $\sqrt{3} \times \sqrt{3}$ ML silicene is hard to be synthesized, being consistent with the previous experimental results \cite{padova,resta}.
As for the $\pi$-bonded chain structure, commonly observed on Si(111) surface, the topmost two surface layers show drastic reconstruction. Hence the ML silicene cannot evolve to the $\pi$-bonded chain structure.

\subsection{Bilayer Si on Ag(111)}

In this subsection, we present the results for the BL silicene.
We focus on the $3 \times 3$ BL silicene on $4 \times 4 $ Ag(111) surface and on the $\sqrt{7} \times \sqrt{7}$ BL silicene on  $\sqrt{13} \times \sqrt{13}$ Ag(111)
, where the lattice mismatches are
0.4 \% and 2.6 \% (with respect to the (111) plane of diamond-structured Si), respectively. Another degree of freedom is the way of stacking of the two Si layers. When the two planar hexagonal networks are stacked in the same way, we call it the AA stacking, whereas when the two networks are rotated
by 60 degrees we call it the AB stacking, following the convention of graphite. From these two different stacking ways, we start the geometry optimization. We also examine stability of the $\pi$-bonded chain structure
and the DAS structure
for this BL Si atomic layers on the Ag(111) surface. One of the interesting features is the multi-stability of various structures of which the cohesive energies are in the range of 5.316-5.350 eV.

\begin{figure}
\begin{center}
\includegraphics[angle= 0,width=0.8\linewidth]{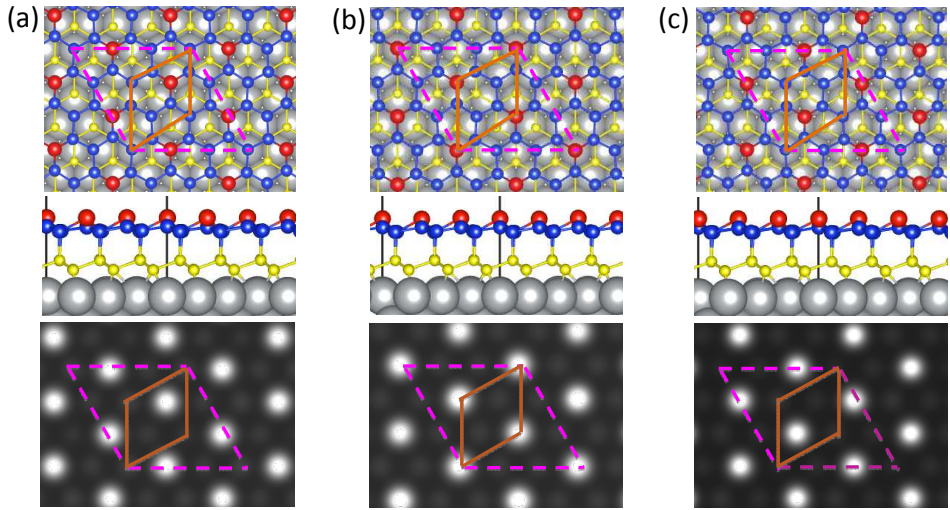}
\caption{
(color online)
Top (upper panels) and side (middle panels) views of the calculated stable structures and corresponding STM images (bottom panels) of the three rhombic $\sqrt{3} \times \sqrt{3}$ bilayer (BL) silicene on Ag(111). The (a), (b) and (c) are
labeled as str1, str2 and str3 of BL silicene, respectively. The large gray balls and the small yellow balls depict the positions of Ag atoms of substrate and the Si atoms of the bottom Si layer, respectively. The large red balls and small blue balls depict the positions of protruded and unprotruded Si atoms of the top Si layer.
The lateral unit cells are indicated by the dashed (pink) lines in the top views, and their boundaries are depicted by the solid (black) lines in the side views.
The obtained rhombic $\sqrt{3}\times\sqrt{3}$ periodicity is indicated by the solid (orange) lines.
}
\label{BLs1}
\end{center}
\end{figure}

Starting from the $3 \times 3$ BL silicene on the $4\times4$ Ag(111) with AB stacking, we have reached three stable $\sqrt{3} \times \sqrt{3}$ rhombic silicene structures shown in Fig.\ref{BLs1}. The three structures, labeled as str1, str2 and str3, are equivalent to each other when we neglect the existence of the Ag substrate:
In the
top layer, a single Si atom is protruded by about 1 {\AA} from the remaining five Si atoms in the lateral $\sqrt{3}\times\sqrt{3}$ periodicity, similar to the $\sqrt{3}\times\sqrt{3}$ structure of ML silicene in Fig. \ref{strSL}(a). In the bottom silicene layer, Si atoms are buckled with the amount of about 0.8 {\AA}, as they are on the (111) plane of diamond-structured Si. These features are common for the three structures. Moreover, the total-energy difference among the three structures are within 0.36 meV/Si, inferring coexistence of all three structures in usual experiments.
These structures can be transformed to each other by moving the rhombic $\sqrt{3}\times\sqrt{3}$ unit cells along their diagonal direction.

The calculated STM images \cite{stm} of the three $\sqrt{3}\times\sqrt{3}$ structures show the perfect rhombic geometry, which agrees well with that observed by Chen et al. in experiment below 40K \cite{chen2}. As will be discussed in section 4, the flip-flop motion would happen between two of the three structures at higher temperatures. As a result, the honeycomb $\sqrt{3}\times\sqrt{3}$ STM image [Fig. 8(a)] that is widely reported in the experiments \cite{padova,APL,2DM} will be obtained.

\begin{figure}
\begin{flushright}
\includegraphics[angle= 0,width=0.9\linewidth]{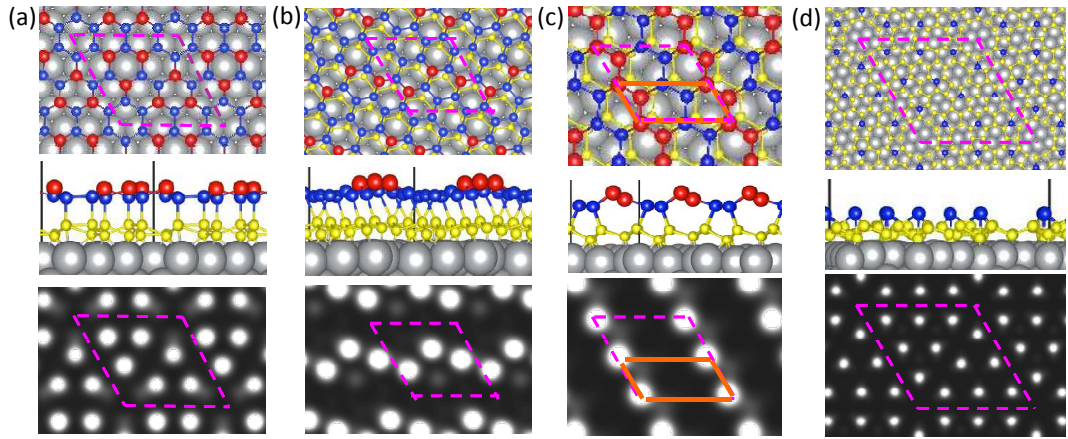}
\caption{
(color online)
Top (upper panels) and side (middle panels) views of the calculated stable structures and corresponding STM images (bottom panels) of BL silicene and the $\pi$-bonded chain and the DAS structures on Ag(111). (a), (b) The $3\times3$ and the $\sqrt{7} \times \sqrt{7}$ BL silicene structures, respectively. (c), (d) The $2\times1$ and the $7\times7$ BL Si(111)-surface structures, respectively.
The color codes are the same as in Fig.~\ref{BLs1}. The obtained $2\times1$ periodicity in (c) is indicated by the solid (orange) lines.
}
\label{BLs2}
\end{flushright}
\end{figure}

The structure of the $3 \times 3$ BL silicene on the $4\times4$ Ag(111) with AA stacking keeps its periodicity even after the geometry optimization [Fig.\ref{BLs2}(a)].
The top layer of the structure resembles the $3 \times 3$ ML silicene on Ag(111) [Fig. \ref{strSL}(b)].
The corresponding STM image is further calculated , where each protruded Si atom in the top layer corresponds to a bright spot [bottom panel of Fig. \ref{BLs2}(a)]. The STM image is nearly the same as that of the $3\times3$ ML silicene on Ag(111) which has been extensively observed in experiments \cite{vogt}. It is known that the usual experimental approaches (STM and LEED) can only reflect the geometry feature of the top silicene
layer.
This means that people are easily mislead by the STM/LEED geometries in identifying the thickness of silicene in experiments. Although, the height along particular direction of the silicene layers can be measured and thus used to distinguish silicene thickness, it does not work when the Ag surface is not well verified \cite{Lay} or the Ag surface is not flat\cite{resta}.
This indicates that not only the STM measurement but also the determination of the Si coverage is important for the structural identification.

As for the $\sqrt{7} \times \sqrt{7}$ BL silicene on $\sqrt{13} \times \sqrt{13}$ Ag(111), we have also considered both
AA and AB stacking cases. The most stable structure we have found is the $\sqrt{7} \times \sqrt{7}$ BL silicene with AA$^{\prime}$  stacking [Fig. \ref{BLs2}(b)],
where A$^{\prime}$ means the in-plane dislodgment of the top Si layer from the proper A stacking. The calculated STM image [bottom panel of Fig. \ref{BLs2}(b)] further shows the perfect $\sqrt{7} \times \sqrt{7}$ periodicity with respect to that of $1\times1$ silicene. The most stable AB-stacking and AA-stacking structures we have obtained present the $1\times1$ and $\sqrt{7} \times \sqrt{7}$ periodicities, which are 1.5 meV/Si and 29 meV/Si higher in energy than the AA$^{\prime}$-stacking structure, respectively.

The cohesive energies for the BL silicene on Ag(111) are shown in Table \ref{str}, all of which are larger than that of freestanding silicene (4.75 eV/Si in our calculation), showing the catalytic function of Ag in forming silicene.
On the other hand, the cohesive energies of BL silicene on Ag(111) are smaller than those of the ML silicene on Ag(111) by 37-94 meV/Si. This explains why the ML silicene structures are
usually synthesized before the BL silicene in experiments \cite{arafune,padova}. The $3 \times 3$ BL silicene with AA stacking on Ag(111) [$E_c$=5.342 eV/Si, Fig. \ref{BLs2}(a)] is the most stable among all the BL silicene structures, which is lower in the total energy by 3 meV/Si than the $\sqrt{3} \times \sqrt{3}$ silicene structures (AB stacking, Fig. \ref{BLs1}) widely observed in experiments \cite{chen2,arafune,padova,resta}. Considering that the $3\times3$ BL silicene exhibits nearly the same STM image as the $3\times3$ ML silicene on Ag(111) \cite{vogt}, this strongly indicates the existence of the $3\times3$ BL silicene structure on the Ag surface, which may be misrecognized to be the ML silicene in previous experiments. In addition, the $\sqrt{7} \times \sqrt{7}$ BL silicene on Ag(111)
is higher in the total energy by about 10 meV/Si than the $3\times3$ and $\sqrt{3}\times\sqrt{3}$ structures, showing that it is rarely synthesized in experiment.

To explore the possibility of Si(111) reconstructed surface emerging on the Ag(111) substrate,
we deposit the topmost two surface layers of the $\pi$-bonded chain structure \cite{pandey} and the DAS structure \cite{takayanagi,takayanagi2} on Ag(111),
though the two top-layer atoms in the DAS structure is incomplete for the full reconstruction in the DAS model. After geometry optimization, it is found that both structures are stable [Figs. \ref{BLs2}(c) and \ref{BLs2}(d)]. The obtained stable $2\times2$ Si surface shows the periodicity of $2\times1$ in which the bond network in the Si double layer contains five-membered and seven-membered rings and thus two top surface atoms in a unit cell form a $\pi$-bonded chain, being essentially identical to the $\pi$-bonded structure on the Si(111) surface.
We have also calculated the corresponding STM image. As shown in the bottom panel of Fig. \ref{BLs2}(c), the STM image agrees well with that of the pure $2\times1$ Si(111) surface \cite{haneman1}. These results clearly show the emergence of the $2\times1$ Si(111)-surface structure on Ag(111).

The $7\times7$ Si(111)-surface structure is also preserved when we deposit the top two Si layers on Ag(111) [Fig. \ref{BLs2}(d)]: the top layer composed of the 12 Si atoms in a unit cell keeps nearly the same geometry as that of the pure $7\times7$ Si(111) surface, and the geometry of its underlying layer composed of the remaining 90 Si atoms is also basically preserved on Ag surface. The calculated STM image [bottom panel of Fig. \ref{BLs2}(d)] is similar to that of the pure $7\times7$ Si(111) surface \cite{haneman2}. The above results clearly show that the bottom Si layer acts as a buffer layer on the Ag surfaces, which preserves the structure of the Si layers above it.

We have also calculated the cohesive energies for the Si(111)-BL system deposited on Ag(111). It is found that the $\pi$-bonded chain structure on the Ag(111) is the most stable among all the BL Si on Ag(111). The cohesive energy of the $\pi$-bonded chain structure
($E_c$=5.35 eV/Si) is larger
by about 10 meV/Si than those of the $3\times3$ ($E_c$=5.342 eV/Si) and the $\sqrt{3}\times\sqrt{3}$ ($E_c$=5.339 eV/Si) BL silicene structures (Table \ref{str}). As we will explain below, this difference in the cohesive-energy mainly come from the difference in the stability of the top
Si
layer.

A common feature for these BL Si structures (except for the $7\times7$ DAS structure) on Ag(111) is that half amount of Si atoms in the bottom layer located
in the lower positions are chemically bonded with the Ag atoms, whereas the remaining half amount of Si atoms located
at the higher positions are chemically bonded with the Si atoms in top layer (side views of Figs. \ref{BLs1} and \ref{BLs2}). This makes the bottom layer of Si being $sp^3$ bonded with the Ag surface atoms and the top-layer Si atoms. If we consider the Ag substrate and the bottom Si layer as a united system, it can be expected that the cohesive energies of this united systems are nearly the same for various BL Si structures on Ag(111). Therefore, the cohesive-energy difference mainly comes from the surface-energy difference of the
top layer
for all the BL Si structures on Ag(111).
To confirm this, we have explored the stable structures of pure Si(111) surface with the $\sqrt{3} \times \sqrt{3}$ periodicity and found that its cohesive energy is $E_c$=5.18 eV/Si, being smaller than the pure $2 \times 1$ $\pi$-bonded Si(111) surface by about 10
meV/Si. This amount coincides with the cohesive-energy difference between the $\sqrt{3}\times\sqrt{3}$ and $2\times1$ BL Si structures on Ag(111).

It should be mentioned that the $7\times7$ BL Si(111)-surface structure does not follow the rule stated above.
The $7 \times 7$ DAS structure is most stable on Si(111) surface, whereas it is least stable as the BL system on the Ag(111) (Table \ref{str}). This is due to the complex reconstruction in the DAS model in which subsurface atoms are relaxed substantially. The BL system on the Ag(111) is not thick enough to incorporate the complex reconstruction.
However, we expect that the $7\times7$ Si(111)-surface structure on Ag(111) would be the most stable when the number of Si layers increases. Yet this issue will not be discussed in this paper mainly due to the limited resource of the computation.

\subsection{Tri-layer Si on Ag(111)}
\begin{figure}
\begin{center}
\includegraphics[angle= 0,width=0.8\linewidth]{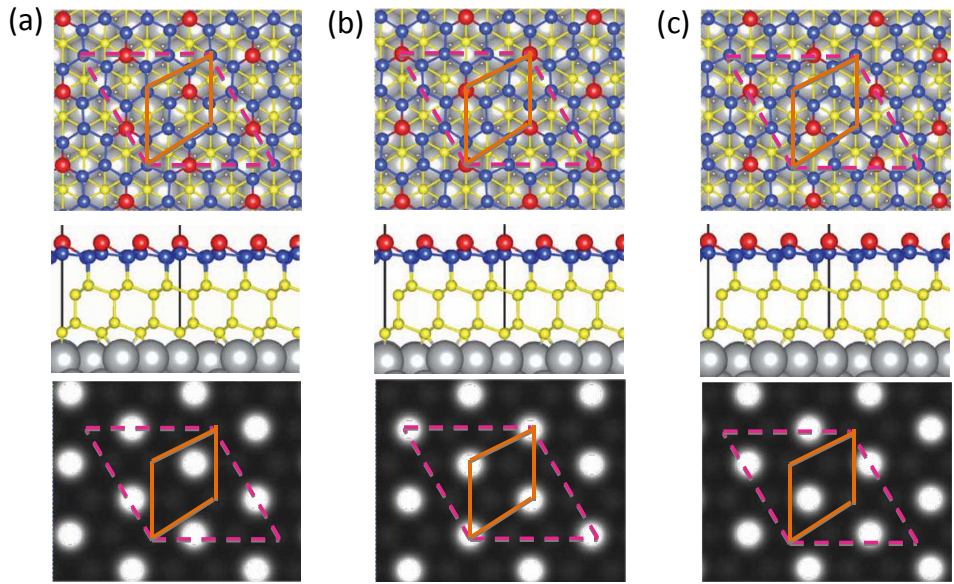}
\caption{
(color online)
Top (upper panels) and side (middle panels) views of the calculated stable structures and corresponding STM images (bottom panels) of the three rhombic $\sqrt{3} \times \sqrt{3}$ tri-layer (TL) silicene on Ag(111). The (a), (b) and (c) are
labeled as str1, str2 and str3 of TL silicene, respectively. The large gray balls and the small yellow balls depict the positions of Ag atoms of substrate and the Si atoms under the top Si layer, respectively. The large red balls and small blue balls depict the positions of protruded and unprotruded Si atoms of the top Si layer.
The simulated lateral unit cell is indicated by the dashed (pink) lines in the top view, and its boundary is indicated by the solid (black) lines in the side views. The obtained rhombic $\sqrt{3}\times\sqrt{3}$ periodicity is indicated by the solid (orange) lines.
}
\label{TLs1}
\end{center}
\end{figure}

\begin{figure}
\begin{flushright}
\includegraphics[angle= 0,width=0.9\linewidth]{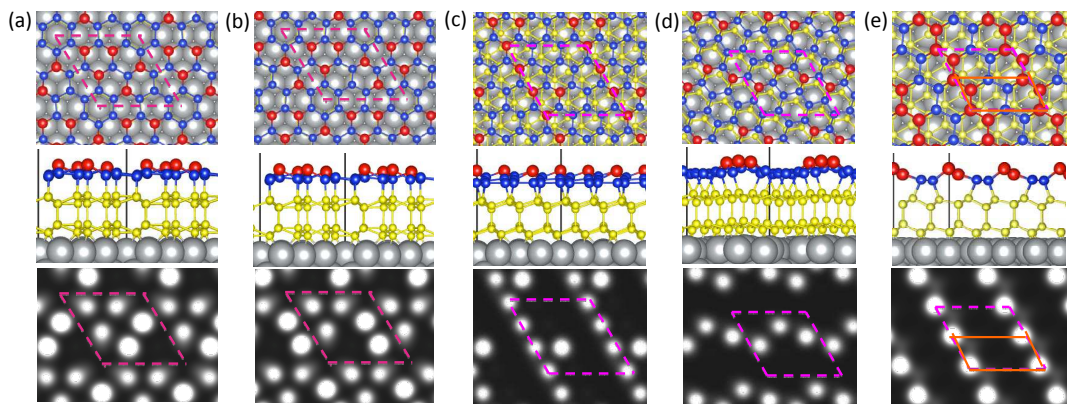}
\caption{
(color online)
Top (upper panels) and side (middle panels) views of the calculated stable structures and corresponding STM images (bottom panels) of the TL silicene and Si(111)-surface structures on Ag(111). (a), (b) The two
symmetry-related $3 \times 3$ silicene structures with the AAA stacking, which are labeled
as str1 and str2, respectively. (c) The $3\times3$ silicene structure with ABC$^{\prime}$ stacking. (d) The $\sqrt{7}\times\sqrt{7}$ silicene structure. (e) The $2\times1$ Si(111) $\pi$-bonded chain structure. The color codes are the same as in Fig.~\ref{TLs1}.
The obtained $2\times1$ periodicity in (e) is indicated by the solid (orange) lines.
}
\label{TLs2}
\end{flushright}
\end{figure}

In this subsection, we present the results for the TL silicene.
Let us start with $3 \times 3$ TL silicene on $4\times4$ Ag(111). Interestingly, we have found three
stable
$\sqrt{3} \times \sqrt{3}$ rhombic silicene structures for the TL silicene [Fig. \ref{TLs1}], as in the $3 \times 3$ BL silicene.
The three structures are transformed to each other and the total energies of the three structures are nearly the same, showing the tristability of the $\sqrt{3} \times \sqrt{3}$ TL silicene.
The three Si layers are stacked as ABC, being identical to the stacking of the Si(111).
The structure of the top silicene layer of the TL silicene is very similar to that in the BL $\sqrt{3} \times \sqrt{3}$ silicene. The calculated STM images are nearly the same with those of the $\sqrt{3} \times \sqrt{3}$ BL silicene on Ag(111) [the bottom panels of Fig. \ref{TLs1}].

Since we have found the $3\times3$ AA-stacking BL silicene [Fig. \ref{BLs2}(a)], we explore the similar $3\times3$ TL silicene structure with AAA stacking where the top layer resembles that of the AA-stacking BL silicene. As a result, the $3 \times 3$ structure is found to be unstable and relaxes into either of other two $3 \times 3$ structures with AAA stacking [str1, str2 shown in Figs. \ref{TLs2}(a) and \ref{TLs2}(b)], where 5, not 6 in the case of the BL silicene, of 18 Si toms in the top layer protrude by about 0.9 {\AA}. The resultant two structures are symmetrically equivalent if we neglect the existence of the Ag substrate. The total energy difference between the two
symmetry-related structures is therefore nearly zero, showing the bistability of the $3\times3$ structures in TL silicene. The calculated STM images of the two structures reflect the characteristics of the top-layer atom arrangements [bottom panels of Figs. \ref{TLs2}(a) and \ref{TLs2}(b)].

We have found a new $3 \times 3$ TL silicene structure where the stacking of the three Si layers are dislodged [Fig. \ref{TLs2}(c)]: The stacking is ABC$^{\prime}$ where the top layer C is dislodged in the lateral plane from the proper C stacking as
in diamond-structured Si. In the top layer, 4 of 18 Si atoms are protruded by about 1 {\AA}, whereas the remaining 14 Si atoms locate in the lower positions. This causes very different STM profile from the other Si structures on Ag(111) [bottom panel of Fig. \ref{TLs2}(c)]. Since the cohesive energy of this structure is large enough (see below), this unusual STM pattern should be observed.

As for the $\sqrt{7} \times \sqrt{7}$ TL silicene on
$\sqrt{13}\times\sqrt{13}$ Ag(111), we have found
that the most stable
structure keeps the $\sqrt{7} \times \sqrt{7}$ periodicity with the AAA$^{\prime}$ stacking [Fig. \ref{TLs2} (d)]. The structure of top silicene layer and the STM pattern are nearly the same as those of the $\sqrt{7} \times \sqrt{7}$ BL silicene [Fig. \ref{BLs2} (b)].

The most prominent characteristic of the stable ML, BL, and TL silicene obtained in the present calculations is the protrusion of some of top-layer Si atoms causing the nearby buckling of the top layer. This structural relaxation was originally proposed to explain $ 2 \times 1$ Si(111) surface by Haneman \cite{haneman3} before the $\pi$-bonded chain model was proposed by Pandey \cite{pandey}. Hence it is recognized that the restriction in the lateral plane caused by the Ag substrate realizes the buckling relaxation which is irrelevant in the free Si(111 ) surface. The buckling structure caused by the external environment may induce unexpected phenomena such as spin polarization \cite{okada}.

To verify the stability of the $\pi$-bonded structure in the TL silicene on the Ag(111),
we deposit the topmost three surface layers of the $\pi$-bonded chain structure
on Ag(111).
After geometry optimization, it is found that the $\pi$-bonded chain structure is stable [Figs. \ref{TLs2}(e)].
The calculated STM image [bottom panel of Fig. \ref{TLs2}(e)] agrees well with that of the $2\times1$ $\pi$-bonded chain structure on the Si(111) surface \cite{haneman1}.

The calculated cohesive energies for the TL Si structures on Ag(111) are shown in Table \ref{str}. A common feature is that the cohesive energies in the TL structures are obviously larger than those of BL structures by tens of meV/Si, indicating that it is easier to obtain the TL Si structures on Ag(111). The most stable silicene structure is the $3\times3$ silicene with ABC$^{\prime}$ stacking [$E_{c}$=5.391 eV/Si, Fig. \ref{TLs2}(c)]. This is different from that in the BL silicene, where the $3\times3$ silicene with AA stacking is the most stable.
The cohesive
energy of the $\sqrt{3}\times \sqrt{3}$ TL silicene structure ($E_{c}$=5.384 eV/Si) is very close to that of the most stable TL silicene structure,
indicative of their coexistence.
On the other hand, the $3\times3$ and $\sqrt{7}\times \sqrt{7}$ silicene structures with AAA and AAA$^{\prime}$ stacking are smaller in $E_c$ by 20-26 meV/Si than the most stable one, indicating that it may be difficult to observe them in experiments.

It is noteworthy that the $2\times1$ $\pi$-bonded chain structure is stable even on the Ag(111) substrate and further its cohesive energy ($E_c$=5.403 eV/Si) is larger than that of the most stable silicene structure ($E_c$=5.391 eV/Si) by about 10 meV/Si. This confirms our
argument stated above:
The relaxed or reconstructed Si(111)-surface structures on Ag(111) are more stable than the silicene structures owning to their lower surface energies.

\subsection{Quad-layer Si on Ag(111)}
\begin{figure}
\begin{center}
\includegraphics[angle= 0,width=0.8\linewidth]{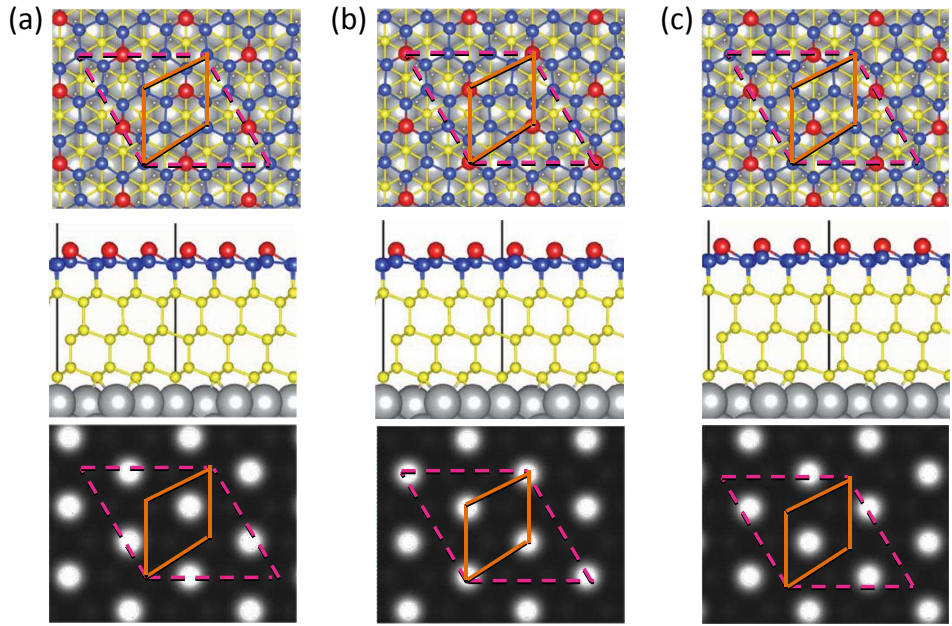}
\caption{
(color online) Top (upper panels) and side (middle panels) views of the calculated stable structures and corresponding STM images (bottom panels) of the three rhombic $\sqrt{3}\times\sqrt{3}$ quad-layer (QL) silicene on Ag(111)
labeled as str1 (a), str2 (b) and str3 (c), respectively. The large gray balls and the small yellow balls depict the positions of Ag atoms of the substrate and the Si atoms under the top Si layer, respectively. The large red balls and small blue balls depict the positions of the protruded and unprotruded Si atoms of the top Si layer.
The simulated lateral unit cell is indicated by the dashed (pink) lines in the top view, and its boundary is indicated by the solid (black) lines in the side views.
The obtained rhombic $\sqrt{3}\times\sqrt{3}$ periodicity is indicated by the solid (orange) lines.
}
\label{QLs1}
\end{center}
\end{figure}

\begin{figure}
\begin{flushright}
\includegraphics[angle= 0,width=0.9\linewidth]{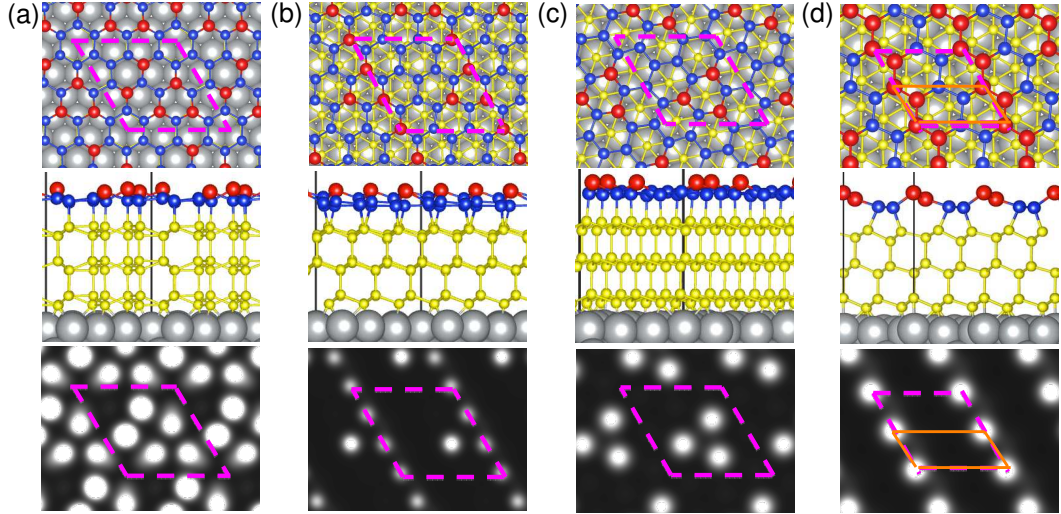}
\caption{
(color online) Top (upper panels) and side (middle panels) views of the calculated stable structures and corresponding STM images (bottom panels) of the QL silicene and the $\pi$-bonded chain structures on Ag(111). (a), (b) The $3\times3$ silicene structures with AAAA and
ABCA$^{\prime}$ stacking, respectively. (c) The $\sqrt{7}\times\sqrt{7}$ silicene structure with ABCA stacking. (d) The $2\times1$
$\pi$-bonded chain structure. The color codes are the same as in Fig.~\ref{QLs1}.
The obtained $2\times1$ periodicity in (d) is indicated by the solid (orange) lines.
}
\label{QLs2}
\end{flushright}
\end{figure}

In this subsection, we present the results for the QL silicene.
Starting from the $3 \times 3$ TL silicene on the $4 \times 4 $ Ag(111) surface, we have reached the three stable $\sqrt{3}\times\sqrt{3}$ rhombic silicene structures, as in the BL and TL cases.
The stacking is ABCA [Fig. \ref{QLs1}] which seems to be a natural extension of the TL silicene with the ABC stacking.
The structure of the top silicene layer and the STM pattern are nearly the same with those of the BL [Fig. \ref{BLs1}] and TL [Fig. \ref{TLs1}] $\sqrt{3}\times\sqrt{3}$ silicene on Ag(111).

The $3\times3$ silicene structures of AAAA [Fig. \ref{QLs2}(a)] and ABCA$^{\prime}$ [Fig. \ref{QLs2}(b)] stackings are also
stable.
The AAAA structure seems to be an extension of the structures of TL silicene with the AAA stacking [Figs. \ref{TLs2}(a) and \ref{TLs2}(b)]. While the top-layer structure and the STM pattern present a different rotational angle with respect to the Ag surface. Other structure mapped by the symmetry operations is not found for the QL case.
In the ABCA$^{\prime}$ structure, the bottom 3 Si layers present the perfect ABC stacking.
The top Si layer (A$^{\prime}$ layer) presents a in-plane dislodgment from the proper A stacking, the geometry of which resembles the C$^{\prime}$ layer of ABC$^{\prime}$-stacking structure in the TL case, but with a different rotational angle with respect to the Ag surface. Thus the STM pattern [bottom pattern of Figs. \ref{QLs2}(b)] presents some similarity with that of the ABC$^{\prime}$ TL silicene on Ag(111) [bottom pattern of Figs. \ref{TLs2}(c)].

The most stable structure of the $\sqrt{7}\times\sqrt{7}$ silicene on
$\sqrt{13}\times\sqrt{13}$ Ag(111) also presents the $\sqrt{7}\times\sqrt{7}$ silicene periodicity [ABCA stacking, Fig. \ref{QLs2}(c)] as that in the TL case [Fig. \ref{TLs2}(d)], where 3 of 14 Si atoms in the top layer protruded by about 1 {\AA} constitute the triangle geometry.
On the other hand,
the STM images of the two $\sqrt{7}\times\sqrt{7}$ structures are different from each other [bottom panels of Fig. \ref{QLs2}(c) and Fig. \ref{TLs2}(d)] due to the different positions of the protruded Si atoms. This shows the Si thickness dependence of the most stable silicene structures for the $\sqrt{7}\times\sqrt{7}$ silicene on $\sqrt{13}\times\sqrt{13}$ Ag(111).

To verify the stability of the  $2 \times 1$ $\pi$-bonded structure in the QL silicene on the Ag(111),
we deposit the topmost four surface layers of the $\pi$-bonded chain structure
on Ag(111) and perform the geometry optimization. As a result, the $\pi$-bonded chain structure is also stable [Fig. \ref{QLs2}(d)] on Ag(111) and the calculated STM image [bottom panel of Fig. \ref{QLs2}(d)] agrees well with that of the $2\times1$ $\pi$-bonded chain structure on the Si(111) surface \cite{haneman1}.

The calculated cohesive energies for the QL Si structures on Ag(111) are shown in Table \ref{str}. The cohesive energies in the QL structures are obviously larger than those of the BL and TL structures, showing the increment of cohesive energies with the number of layers for multilayer silicene.
The most stable silicene structure is the $3\times3$ structure with ABCA$^{\prime}$ stacking [$E_c$=5.423 eV/Si, Fig. \ref{QLs2}(b)]. The cohesive energy of the $\sqrt{3}\times\sqrt{3}$ silicene structure [$E_c$=5.418 eV/Si] is very close to it, showing their coexistence.
The $\sqrt{7}\times \sqrt{7}$ and $3\times3$ silicene structures with ABCA and AAAA stackings are smaller in $E_c$ by 19-39 meV/Si than the most stable one, implying that it is hard to synthesize them in experiments.
The cohesive energy ($E_c$=5.434 eV/Si) of $2\times1$ $\pi$-bonded chain structure is larger than that of the most stable silicene structure also by about 10 meV/Si, confirming that the relaxed or reconstructed Si(111)-surface structures are more stable than the silicene structures on Ag(111).

The reason why the $2\times1$ $\pi$-bonded chain and/or $7\times7$-DAS structures have not been observed on Ag(111) in experiments is attributed to the synthesizing method, where the
structures are always produced via depositing the Si atoms on Ag(111). This makes the Si structures on Ag(111) grown layer by layer: the BL Si structures are grown based on the structures of ML
silicene and the TL Si structures are on the BL Si structures etc.
Since the structure of ML silicene is very different from that of the Si(111) layer, it is not expected to obtain the $2\times1$ BL Si(111)-surface structure on Ag(111) by using the synthesizing method widely adopted to make silicene. It is similar for synthesizing thicker (TL, QL and so forth) Si structures on Ag(111). On the other hand, it is well known that the $2\times1$ Si(111) structure is obtained by cleaving the Si(111) surface at room temperature, and it further converts to the more stable $7\times7$ structure when heated above $400~^{\circ}C$. This shows that different experimental approaches produce the different Si structures.
Therefore, although the $2\times1$ and $7\times7$ Si structures are more stable than the silicene structures on Ag(111), they have never been observed in experiment because
no one
has performed such experiments as
heating the multilayer silicene structures at high temperatures to convert the silicene structures to the $2\times1$ or $7\times7$ Si(111) structures.

Table \ref{str} also shows the structure details and the binding energies for all the Si structures from ML to QL on Ag(111). The binding energies for all the structures (the energy gain in the deposition of Si layers on the Ag surface) are in the range of 0.6-0.9 eV/Si, larger than the typical vdW interaction energy manifested in the cases of graphene on metal surfaces by an order of magnitude \cite{Vanin}, showing the covalent nature of the Si-Ag interface interactions. Moreover, the spacings between the bottom Si layer and Ag surface are in the range of 2.17-2.37 {\AA}. Considering that the atomic radii of Si and Ag are 1.18 and 1.65 {\AA}, respectively, the small spacing values confirm the formation of the Si-Ag covalent bonds.
This result agrees with the recent DFT results which studied various monolayer silicene structures on Ag(111) \cite{pflugradt, kaltsas}. While it disagrees with the argument in Ref. \cite{scalise} that the Si-Ag interaction is weak.
On the other hand, the interlayer distances between the top Si layer and the underlying layer for all the structures are smaller than 2.6 {\AA}. This shows the covalent nature of Si-Si interlayer interactions. As discussed below, the covalent interactions of Si-Ag interface and Si-Si interlayer would significantly influence the electronic properties of silicene.

\section{Flip-flop motion in multilayer silicene on Ag(111)}
\begin{figure}
\begin{flushright}
\includegraphics[angle= 0,width=0.9\linewidth]{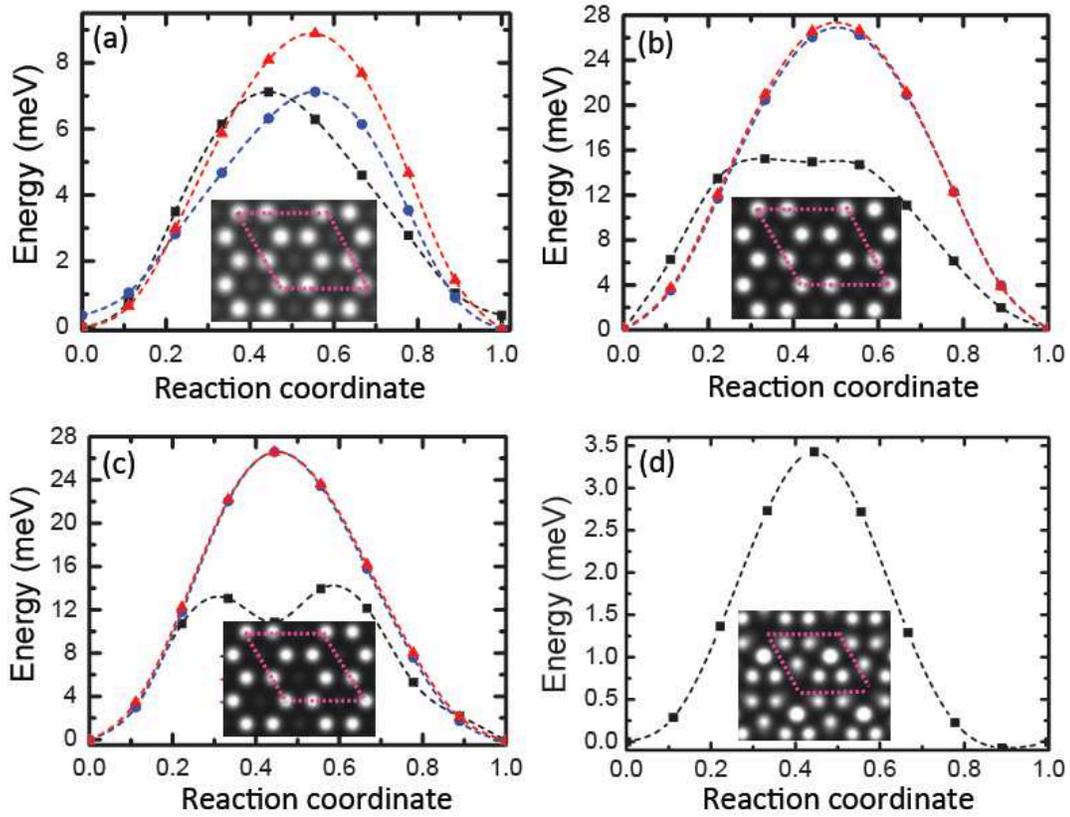}
\caption{(color online) Calculated reaction energy profile among the three rhombic $\sqrt{3}\times\sqrt{3}$ structures of BL (a), TL (b) and QL (c) silicene on Ag(111), shown in Figs. \ref{BLs1}, \ref{TLs1} and \ref{QLs1}, respectively. (d) The calculated reaction energy profile between the two rotation-symmetric $3\times3$ structures of TL silicene with AAA stacking on Ag(111) [Figs. \ref{TLs2}(a) and \ref{TLs2}(b)].
The energy in the vertical axis is defined as $(E-E_{str1})/n$, where E and $E_{str1}$ are the total energy of the structure at reaction coordinate and the total energy of str1, respectively. $n$ is the number of atoms in the top silicene layer. The transition barriers between str1 and str2, str2 and str3, and str1 and str3 are depicted by the black squares, blue dots and red triangles, respectively.
The insets show the STM images obtained by superimposing the images of str1 and str2 structures. The simulated $3\times3$ lateral unit cells in the top views are indicated by the dashed (pink) lines.

}
\label{neb}
\end{flushright}
\end{figure}

In this section, we present the flip-flop motion of multilayer silicene on Ag surface. We focus on the $\sqrt{3}\times\sqrt{3}$ tristable BL, TL and QL silicene structures and the $3\times3$ bistable TL silicene structures. The calculated transition barriers are indicative of the flip-flop motion among/between the tristable/bistable silicene structures at low temperature. The results agree well with the STM images observed in recent experiments.

In use of the nudged elastic band (NEB) method \cite{neb1,neb2}, we have calculated the transition barriers among the three $\sqrt{3}\times\sqrt{3}$ structures, str1, str2, str3 of each BL, TL and QL silicene on Ag(111) [Figs.~\ref{neb}(a), \ref{neb}(b) and \ref{neb}(c)]. The calculated transition barriers for the BL silicene are in the range of 7-9 meV/Si, whereas they are in the range of 14-28 meV/Si for the TL and QL silicene on Ag(111). Anyhow, these barriers are low enough to allow the
structural transition to be thermoactivated at temperatures of dozens of Kelvin.
The insets of Figs. \ref{neb}(a), \ref{neb}(b) and \ref{neb}(c) show superpositions of two of the three STM images for the BL, TL, and QL silicene structures, respectively, which present the perfect honeycomb $\sqrt{3}\times\sqrt{3}$ structures. These results agree well with the recent experimental observations
\cite{chen2,padova,resta,APL,2DM,padova1}.
It is noted that two other models had been recently proposed to explain the $\sqrt{3}\times\sqrt{3}$ bilayer silicene structure observed in experiments \cite{pflugradt1,cahangirov1}. However, both of them can not explain all the experimental observations at lower (below 40K, rhombic structure)\cite{chen2} and higher temperatures (above 77K, honeycomb structure)
\cite{chen2,padova,resta,APL,2DM,padova1}:
The model in Ref.\cite{pflugradt1} cannot explain the honeycomb STM structure and the model in Ref.\cite{cahangirov1} cannot explain
its conversion from the honeycomb STM structure to the rhombic STM structure at lower temperature \cite{chen2}.
From this sense, our model agrees better with the experimental observations so far.
From the previous section, it is recognized that the bottom Si layer acts as a buffer layer on the Ag surfaces, above which the structure of Si layers can be well preserved. Thus the tristable $\sqrt{3}\times\sqrt{3}$ silicene structures and the low transition barriers among them are also expected to exist in the thicker multilayer silicene on Ag(111). This explains the experimental observations where the same honeycomb $\sqrt{3}\times\sqrt{3}$ STM images are observed independent of the thickness of silicene \cite{chen2,arafune,padova,resta,APL,2DM,padova1}.

The flip-flop motion also happens in other silicene structures. Fig. \ref{neb} (d) shows the calculated transition barriers between the two bistable structures of the $3\times3$ TL silicene with AAA stacking [Figs. \ref{TLs2}(a) and \ref{TLs2}(b)].
The very small energy barrier (3.5 meV/Si) shows that the flip-flop motion can happen at very low temperature.
The inset shows the superposition of the STM images of the two structures, which exhibits the similar honeycomb $3\times3$ STM image as that of the $3\times3$ ML and BL silicene [bottom panel of Fig. \ref{BLs2}(a)] on Ag surfaces. The results confirm the characteristic of flip-flop motion in multilayer silicene on Ag(111) at low temperature.

\section{Dirac states and surface-localized $\pi$ states for the multilayer Si on Ag(111)}
\begin{figure}
\begin{flushright}
\includegraphics[angle= 0,width=0.99\linewidth]{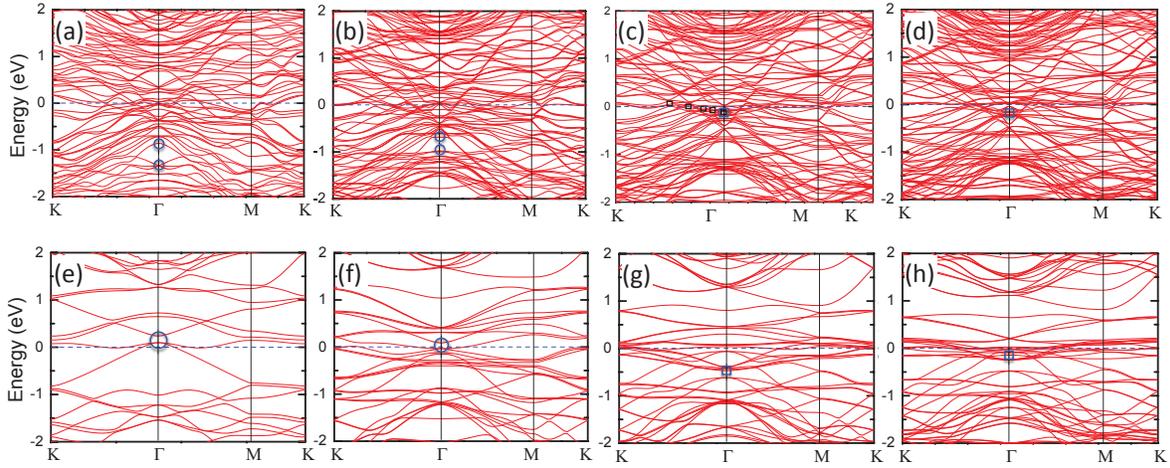}
\caption{
(color online) Calculated energy bands of the $3\times3$ silicene on $4\times4$ Ag(111), where the silicene exhibits the $\sqrt{3}\times\sqrt{3}$ structure.
The energy bands of the ML, BL, TL, and QL silicene on Ag(111) are shown in
(a), (b), (c) and (d).
The energy bands of the freestanding ML, BL, TL, and QL silicene that are peeled from the Ag surface are shown in (e), (f), (g), (h), respectively. The origin of the energy is set to be $E_{\rm F}$.
The states that have characters of the $\pi$ ($\pi^{*}$) for the ML and BL silicene on Ag(111) are indicated
by the blue circles on $\Gamma$ point. The surface-$\pi$ states in
TL and QL silicene
are indicated by the blue squares on $\Gamma$ point. The states of the TL silicene on Ag(111) that have characters of the surface-$\pi$ along the $\Gamma-K$ direction are marked by the sequence of black squares in (c), which present linear energy dispersion.
}
\label{bd}
\end{flushright}
\end{figure}

\begin{figure}
\begin{flushright}
\includegraphics[angle= 0,width=0.99\linewidth]{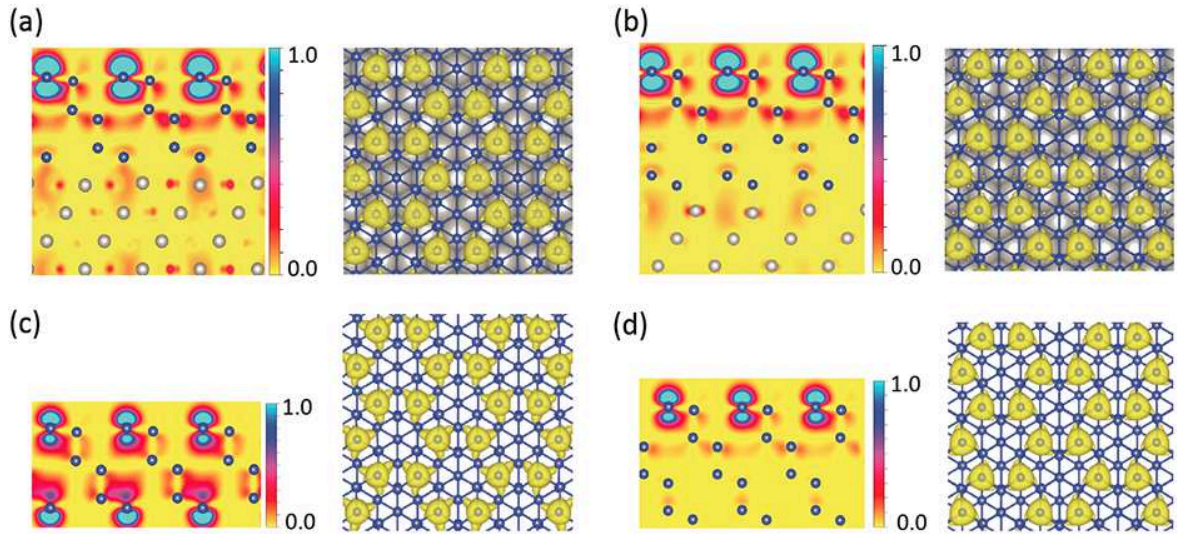}
\caption{
(color online) Contour plots (left panels) and isovalue surface at its value of 10\% of the maximum value (right panels) of the squared Kohn-Sham (KS) orbitals marked by the squares in Figs.~\ref{bd}(c) and (d)
of the $\sqrt{3}\times\sqrt{3}$ TL (a) and QL (b) silicene on Ag(111).
Figures (c) and (d) show the corresponding contour plots and isovalue surface of the KS orbitals marked in  Figs.~\ref{bd}(g) and \ref{bd}(h) of the $\sqrt{3}\times\sqrt{3}$ TL and QL silicene peeled from the Ag surfaces.
The gray and blue balls depict Ag and Si atoms, respectively.
}
\label{KS}
\end{flushright}
\end{figure}

Previously, we have studied the electronic properties of $3\times3$ and $\sqrt{7}\times\sqrt{7}$ ML silicene on Ag(111), which are very different from that of the freestanding silicene \cite{guo1,guo2}. A common feature for all the silicene structures on Ag(111) is
that the Dirac $\pi$ and $\pi^{*}$ states near $E_{F}$ in the freestanding silicene disappear due to
the strong silicene-Ag interactions.
In this section, we present the results for the $\sqrt{3}\times\sqrt{3}$ silicene on Ag(111) because they are the only well observed multilayer silicene structures. Since the $\sqrt{3}\times\sqrt{3}$ structure has stable phases from ML to QL, such study can help us to understand the evolution of electronic properties with the
increasing Si thickness
on Ag surfaces. Additionally, we explore the effect of Ag substrate on the electronic properties of Si(111)-surface structures, i.e., the $2\times1$ $\pi$-bonded chain structure.

The energy bands are calculated in the periodicity of $3\times3$ silicene on $4\times4$ Ag(111).
Figs. \ref{bd}(a) and \ref{bd}(b)
show the calculated energy bands for the ML and BL silicene on Ag(111), where the K point in the BZ
of the $1\times1$ silicene is folded on the $\Gamma$ point.
For both ML and BL cases, we have found no apparent energy band near $E_{F}$ that shows the linear energy dispersion peculiar to Dirac $\pi$ ($\pi^{*}$) state. Instead, the states with characters of $\pi$ ($\pi^{*}$) appear deep below $E_{F}$ [indicated by blue circles in
Figs. \ref{bd}(a) and \ref{bd}(b)]. The energy bands of ML and BL silicene which
we obtain by peeling from the Ag surface are further calculated. It is found that the states with characters of $\pi$ ($\pi^{*}$) shift upward near $E_{F}$ from the deep positions in the valence bands due to the annihilation of
the interaction of Ag substrate [indicated by blue circles in
Figs. \ref{bd}(e) and \ref{bd}(f)].
This shows that the covalent interactions between silicene and Ag surface makes the $\pi$ ($\pi^{*}$) states being deep in the valence bands. We also find no linear energy band peculiar to the Dirac $\pi$ ($\pi^{*}$) state near $E_{F}$ for the TL and QL silicene  on Ag(111) [Figs. \ref{bd}(c) and \ref{bd}(d)]. We have additionally calculated the energy bands of the TL and QL silicene peeled from the Ag surfaces [Figs. \ref{bd}(g) and \ref{bd}(h)]. As a result, no Dirac characteristic $\pi$ ($\pi^{*}$) linear energy band near $E_{F}$ is found. This implies that the covalent Si-Si interlayer interactions also
cause the absence of Dirac states. These results strongly indicate the absence of Dirac states for multilayer silicene on Ag(111) due to the covalent interactions at the Si-Ag interface and in the Si layers (Table.1).

On the other hand, we have found a new state near $E_{F}$ for the TL and QL silicene on Ag surfaces, as indicated by the blue squares on $\Gamma$ point in Figs. \ref{bd}(c) and \ref{bd}(d), which are mainly composed of the $\pi$ orbitals
located at the top silicene layer [Figs. \ref{KS}(a) and \ref{KS}(b)]. Here we call the new state
the surface-$\pi$ state.
The distributions of the Kohn-Sham (KS) orbitals for the surface-$\pi$ state [right panels of Figs. \ref{KS}(a) and \ref{KS}(b)] are similar with those of the Dirac $\pi$ ($\pi^{*}$) orbitals in the freestanding ML silicene, both of which present the hexagonal symmetry. Thus we can also expect that the surface-$\pi$ state exhibits linear energy dispersion near $E_{F}$ \cite{SW}. Through the detailed analysis of the KS orbitals along the $\Gamma$-$K$ direction in BZ for all the energy bands, we have indeed observed such linear energy dispersion nearby $E_{F}$ for the surface-$\pi$ state, as indicated by the black squares in Figs. \ref{bd}(c) for the TL silicene on Ag(111). The Fermi velocity of the surface-$\pi$ state is much smaller than that of the Dirac states in the freestanding ML silicene because of the ($\sqrt{3}$ times) larger distances between the two nearest $\pi$ orbitals which
cause the weaker transfer energy.
Moreover, the KS orbitals of the surface-$\pi$ states are
similar for the TL and the QL silicene on Ag(111) [Figs. \ref{KS}(a) and \ref{KS}(b]. This implies that the surface-$\pi$ state always exists near $E_{F}$ for the $\sqrt{3}\times\sqrt{3}$ multilayer silicene on Ag(111), irrespective of the thickness of silicene. We expect that the surface-$\pi$ state has great potential applications in the field of spin electronics \cite{okada}.

To understand the origin of the surface-$\pi$ state, we have further calculated the energy bands of the $\sqrt{3}\times\sqrt{3}$ TL and QL silicene that are peeled from the Ag surface. As a result, the surface-$\pi$ states near $E_{F}$ are also obtained [indicated by the blue squares in Figs. \ref{bd}(g) and \ref{bd}(h)] , although it contains $\pi$ orbitals on the subsurface
Si layer for the TL silicene [Fig. \ref{KS}(c)]. It is noteworthy that the geometries of the surface-$\pi$ orbitals of the TL and QL silicene peeled from Ag(111) [Figs. \ref{KS}(c) and \ref{KS}(d)] are similar with those of the TL and QL silicene on the Ag(111)
[Figs. \ref{KS}(a) and \ref{KS}(b)].
This shows that the surface-$\pi$ state is originated from the unique $\sqrt{3}\times\sqrt{3}$ Si surface structure. Therefore, one can always expect the appearance of surface-$\pi$ state near $E_{F}$ in the thick $\sqrt{3}\times\sqrt{3}$ multilayer silicene structures.

\begin{figure}
\begin{flushright}
\includegraphics[angle= 0,width=0.99\linewidth]{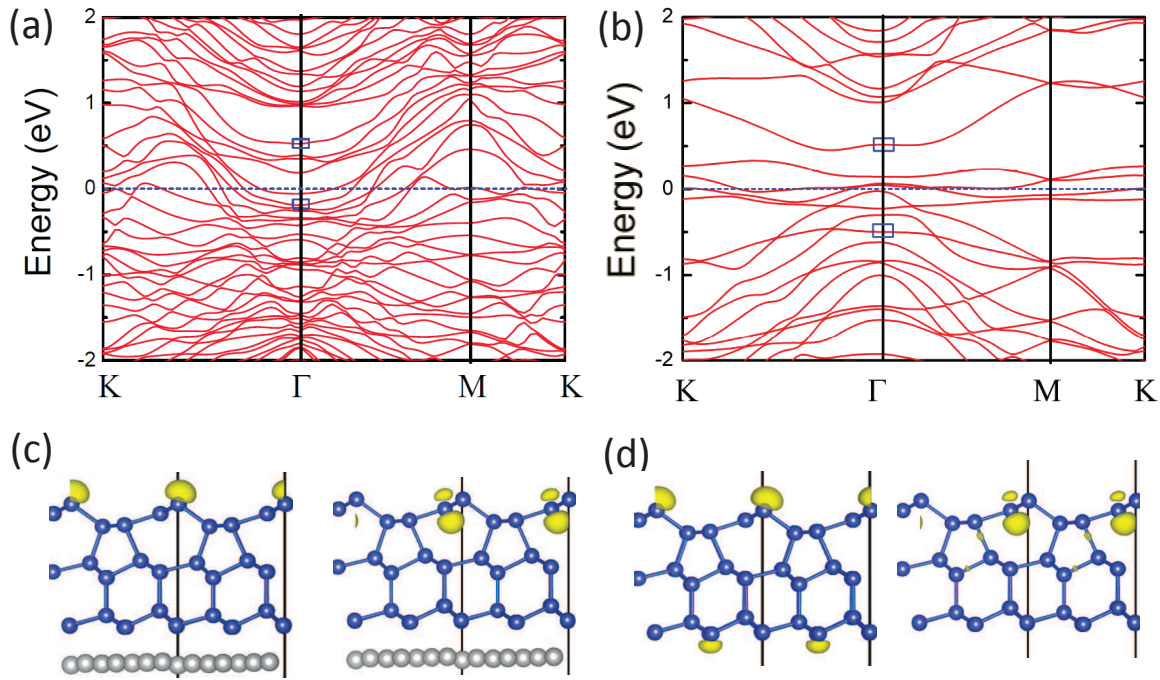}
\caption{
(color online) Calculated energy bands (a) and isovalue surface at its value of 25\% of the maximum value of squared KS orbitals (side view) of the tri-layer $2\times2$ Si(111) on $\sqrt{7}\times\sqrt{7}$ Ag(111), where the Si surface presents $2\times1$ $\pi$-bonded chain structure [Fig. \ref{TLs2}(e)]. The energy bands and isovalue surface of squared KS orbitals of the TL Si structure peeled from Ag surface are further shown in (b) and (d), respectively.
The states that have characters of the isolated $\pi$ ($\pi^{*}$) are indicated the blue squares on $\Gamma$ point in (a) and (b).
The gray and blue balls depict Ag and Si atoms, respectively.
}
\label{2pi}
\end{flushright}
\end{figure}

Since the Si(111)-surface structures are also stable on Ag surfaces, it is important to explore the effects of Ag substrate on the electronic properties of Si(111)-surface structures. It is well known that the $2\times1$ $\pi$-bonded chain Si surface presents a semiconductor property, where both the valence band maximum (VBM) and the conduction band minimum (CBM)
come from the isolated $\pi$ ($\pi^{*}$) states of the $\pi$-bonded atomic chain \cite{zitzlsperger,violante}. Here we have calculated the energy bands of the TL $2\times1$ $\pi$-bonded chain  structure on Ag(111), where the unit cell adopted in the calculation has the periodicity of
$2\times2$ Si(111) on $\sqrt{7}\times\sqrt{7}$ Ag(111)
[Fig. \ref{TLs2}(e)]. The calculated energy bands are shown in Fig. \ref{2pi}(a), which present the metallic property. Through the analysis of calculated KS orbitals, we find the metallic bands are mainly from the sates containing both the Ag orbitals and Si orbitals, as well as the $sp$ band of Ag substrate.
The states with characters of isolated $\pi$ and $\pi^{*}$ appear 0.2 eV below and 0.5 eV above $E_{F}$ on $\Gamma$ point, respectively, as indicated by the blue squares of Fig. \ref{2pi}(a). The corresponding KS orbitals are shown in Fig. \ref{2pi}(c), which clearly shows the isolated $\pi$ and $\pi^{*}$ states
located on the $\pi$-bonded atomic chain.

We have also calculated the energy bands of the $\pi$-bonded chain structure that are peeled from the Ag surface [Fig. \ref{2pi}(b)]. The isolated $\pi$ and $\pi^{*}$ states are found about
0.5 eV below and above $E_{F}$
on $\Gamma$ point [indicated by the blue squares in Fig. \ref{2pi}(b)], respectively. The corresponding KS orbitals are further shown in Fig. \ref{2pi}(d), which resemble those of the $2\times1$ structure on Ag(111), although there are some additional $\pi$ orbitals on the
subsurface
Si layer, as shown in left panel of Fig. \ref{2pi}(d). The energy levels of the isolated
$\pi$ and $\pi^{*}$ states
of the peeled $2\times1$ structure are close to those on the Ag surface, showing the minor effect of the Ag-Si interface interactions on the isolated
$\pi$ and $\pi^{*}$ states
of top Si layer.
We have also obtained the similar results for the QL $\pi$-bonded chain structure on Ag(111). Combining with the results of the $\sqrt{3}\times\sqrt{3}$ silicene on Ag(111), it can be concluded that the strong Ag-Si interface interactions do not destroy the surface-related $\pi$ states of the multilayer Si structures on Ag surfaces.

\section{Conclusion}
We have performed the density-functional calculations for the ultra-thin atomic layers of Si on Ag(111) surfaces, with the thickness of Si layers varying from monolayer to quad-layer.
We have found several stable structures for the silicene on Ag(111) with periodicities of $\sqrt{3}\times\sqrt{3}$, $3\times3$, and $\sqrt{7}\times\sqrt{7}$ with respect to the $1\times1$ silicene. We have also found that the $2\times1$ $\pi$-bonded chain and $7\times7$ DAS Si(111)-surface structures are stable on Ag surfaces, showing the crossover of silicene and Si(111)-surface structures on Ag(111). From the calculated cohesive energies, several common features can be
deduced: (1) ML silicene is more stable than the BL and TL silicene on Ag(111)
but becomes less stable than QL
silicene;
(2) Cohesive energies of multilayer silicene and Si(111)-surface structures on Ag(111) monotonically increase with their thickness increasing; (3) The Si(111)-surface structures are more stable than the silicene structures on Ag(111) due to the lower surface energies.
Moreover, we have found the structural tristability and bistability for the $\sqrt{3}\times\sqrt{3}$ and $3\times 3$ multilayer silicene on Ag(111), respectively. The calculated transition barriers are indicative of the flip-flop motion among/between the tristable/bistable structures at low temperature. The flip-flop motion between two of the three $\sqrt{3}\times\sqrt{3}$ structures produces the honeycomb STM structures observed in experiments.
A common feature for the multilayer Si structures on Ag surfaces is the covalent interactions
at the Si-Ag interfaces and in the Si layers, which cause
the absence of Dirac states in silicene. Instead, we have found a new state with the characters of $\pi$ orbitals
located on the top silicene layer (surface-$\pi$ state), which appears near Fermi level and presents the linear energy dispersion. We expect the surface-$\pi$ states have great potential applications in the spin-related nano-devices.
Finally, we have explored the electronic properties of $2\times1$ $\pi$-bonded chain Si structure on Ag(111). Our results show the robust characteristic of the surface-related $\pi$ states for the multilayer Si structures on Ag surfaces.

\section*{Acknowledgments}
This work was supported by the research project ``Materials Design through Computics" (http://computics-material.jp/index-e.html) by MEXT and also by ''Computational Materials Science Initiative" by MEXT, Japan. Computations were performed mainly at Supercomputer Center in ISSP, University of Tokyo. ZX acknowledges the support of National Natural Science Foundation of China (Grant No. 11204259, 11374252, 11074212, 11275163 and 11304264), the Program for New Century Excellent Talents in University (Grant No. NCET-12-0722), the Program for Changjiang Scholars and Innovative Research Team in University (IRT13093), and the Scientific Research Fund of Education Department of Hunan Province (Grant No. 13B117).


\begin{thebibliography}{10}
\bibitem{ad1}
The optB86b-vdW functional in Ref.\cite{klimes} is adopted in vdW-DF calculations. The calculated lattice constant for the face-centered-cubic Ag is 4.10 {\AA}, which agrees well with experimental value (4.09 {\AA}).
\end{thebibliography}
\end{document}